# Acoustic microstreaming near a plane wall due to a pulsating free/coated bubble: Nyborg's result revisited and vortices

Nima Mobadersany and Kausik Sarkar

Department of Mechanical and Aerospace Engineering, The George Washington University, Washington, DC, USA

**Abstract**

Acoustic microstreaming due to an oscillating microbubble is analytically investigated to obtain the circular streaming motion adjacent to a nearby wall. Classical theory due to Nyborg is carefully derived in the radial coordinates. The theory is used to obtain the flow field and the vortical motion caused by the microbubble motion. The length of the vertices are decreasing when the microbubble is excited at distances close to the rigid wall, while the maximum shear stress is increasing.

**Key words:** sonoporation, microbubble, microstreaming, ultrasound, contrast agent, cavitation, acoustic streaming

## 1. Introduction

One of the interesting characteristics of sound fields is the steady circulation of fluid particles near vibrating elements and bounding walls. It is well known that in addition to the sinusoidal movement in fluid particles, sound sources can generate steady vortices and velocity fields in which particles circulate steadily (Nyborg 1953, Nyborg 1958). These vortex motions have been observed experimentally near oscillating gas bubbles which are resting on a solid surface(Kolb and Nyborg 1956). When associated with microbubbles, this small-scale steady streaming flow is called microstreaming. In this research, we are studying the microstreaming flow near a plane wall analytically.

When the bubble is pulsating, it generates fluctuations in the flow. The time average of these fluctuations usually is not zero creating microstreaming (Riley 2001, Tho, et al. 2007).

Microstreaming has many applications. Microstreaming can be used in micromixing to generate rapid mixing (Liu, et al. 2002, Orbay, et al. 2016). It is shown that bubble induced acoustic micromixing reduces the mixing time for a 22-microliter chamber from several hours (for a pure diffusion-base mixing) to tens of seconds(Liu, et al. 2002). It is also utilized in microfluidic transport to guide solid beads and lipid vesicles in desired direction without microchannels (Marmottant, et al. 2006). Microstreaming can also be used in therapeutic and biomedical applications. The streaming motion of the bubbles exerts shear stress on the boundaries where circulation occurs. It has been shown that the induced shear stress can be exploited in thrombolysis, drug delivery and gene therapy. This shear stress is the main mechanism of hemolysis of the red blood cells (Rooney 1970). Microstreaming can rupture and increase the permeability of cell membranes (sonoporation) in biological tissues for better passage of therapeutic agents across the vascular barrier and cell membranes (Fan, et al. 2014, Marmottant and Hilgenfeldt 2003, Pommella, et al. 2015). Many different streaming patterns is possible due to pulsating bubble

(Collis, et al. 2010, Elder 1959, Tho, et al. 2007). Changing the streaming pattern may result in improved sonoporation and sonothrombolysis (Collis, et al. 2010).

The early theory for microstreaming velocity field and the induced shear stress has been proposed by Nyborg (Nyborg 1958). Rooney used Nyborg's theory approximated for pulsating bubbles resting on a surface hemispherically to calculate maximum shear stress for hemolysis of red blood cells(Rooney 1970). Levin and Bjorno studied the maximum shear stress due to the gas microbubbles on biological cells using Nyborg's theory approximated for hemispherical microbubbles resting on the cells (Lewin and Bjo/rno/ 1982). Forbes and O'Brien determined the theoretical model of Nyborg for estimating microstreaming shear stress with experiments and showed that the theoretical model accurately describes maximum sonoporation activity (Forbes and O'Brien Jr 2012). Doinikov applied Nyborg's theory to find the shear stress on the plane wall by spherical oscillation of contrast agent assuming the bubble is detached from the wall (Doinikov and Bouakaz 2010). Kimmel have applied the axisymmetric boundary element method to calculate the microstreaming shear stress on the plane wall using Nyborg's theory (Krasovitski and Kimmel 2004). Wu used Nyborg's theory to theoretically study the microstreaming shear stress for the Optison contrast agent attached to a wall hemispherical while using Rayleigh-Plesset equation for bubble pulsation (Wu, et al. 2002).

In addition to study the microstreaming shear stress, there are several works studying the microstreaming flow field. Wu and Du solved the streaming generated near both coated and uncoated microbubble in an ultrasound field analytically assuming the wave length is much greater than the bubble radius and showed that the streaming velocity inside the bubble is even much more than the streaming velocity in the outside (Liu and Wu 2009, Wu and Du 1997). Doinikov and Bouakaz calculated acoustic microstreaming near a gas bubble considering all modes of bubble motion without imposing any restriction on the bubble size relative to wave length which would otherwise give rise to underestimation of microstreaming near the bubble (Doinikov and Bouakaz 2010). They also developed a theory for streaming near a bubble in the presence of a distant wall showing that presence of wall gives rise to higher acoustic streaming around the bubble (Doinikov and Bouakaz 2014). Doinikov and Bouakaz described the microstreaming generated by two interacting bubbles and showed that driving the bubbles at near resonance frequencies gives rise to higher microstreaming velocity and stresses (Doinikov and Bouakaz 2016). In addition to theoretical works to study the microstreaming flow field, there are experimental studies showing the flow field around a single and two oscillating bubbles resting on a solid surface (Collis, et al. 2010, Thameem, et al. 2016, Tho, et al. 2007).

Microstreaming flow field near pulsating bubble; uncoated or coated; have both been modeled and experimentally studied. But the microstreaming flow filed near a wall has not been looked in detail while many researches have studied the shear stress induced by microstreaming on a wall. The aim of this study is to show and analyze the microstreaming flow field near a plane rigid wall; assumed as the cell membrane; due to spherical pulsation of microbubbles in an ultrasound field to better understand microstreaming near surfaces. We have shown the streamlines of microstreaming flow near a wall, discussed the reasons for vortex creation, and analyzed the induced shear stress on the wall. We have also studied the length of the vortex and the effect of different parameters on the vortex length.

## 2. Mathematical formulation

The theoretical results on microstreaming (Nyborg 1953, Nyborg 1958) were obtained by solving the governing equations by a perturbative method:

$$\mathbf{u} = \mathbf{u}^{(1)} + \mathbf{u}^{(2)} + ...$$
$$p = p^{(1)} + p^{(2)} + ...$$
(1)

The first order approximation $\mathbf{u}^{(1)}$ solves the linearized equation neglecting the nonlinear advection terms and obtains a sinusoidal velocity. At second order, the convective nonlinear term, quadratic product of $\mathbf{u}^{(1)}$, appear as a forcing term, with the equation upon averaging becomes

$$\mu \nabla^2 <\mathbf{u}^{(2)}>_t - \nabla <p^{(2)}>_t = \mathbf{F} = \rho <\mathbf{u}^{(1)} \cdot \nabla \mathbf{u}^{(1)}>_t$$
(2)

with $\rho$ and $\mu$ being the fluid intensity and viscosity. $<>_t$ is the average over a time period of the oscillating excitation. Averaging the quadratic term gives rise to a steady force that drives the streaming motion. Nyborg noted that the formal solution of this problem requires that boundary conditions be satisfied on the exact boundary, which is possible only for simple boundaries with velocity distributions on them as simple as possible. For a more general situation, Nyborg sought an alternative method. Following an earlier investigation of Schlighting, he made a key observation that one does not need the solution for the entire region (with dimension $L$), but only in the near boundary region (dimension $\delta$). With a number of ingenious approximations, Nyborg was able to obtain an expression of the streaming motion that depends on the surface values of the potential part of the first order velocity. Nyborg's description is terse and based on a formulation with a finite sound speed, and later simplified using "approximate incompressibility" assumption. Furthermore, it was solved in Cartesian coordinates with slight generality for slight curvilinearity. We feel that a brief description of the mathematical derivation in the radial coordinates under the assumption of axisymmetry appropriate for the present problem would be helpful to understand several aspects of the perturbative approach.

*A. Acoustic microstreaming:*

The fluid velocity and pressure $\mathbf{u}(\mathbf{x},t)$ and pressure $p(\mathbf{x},t)$ solve the Navier Stokes equation:

$$\frac{\partial \mathbf{u}}{\partial t} + \mathbf{u} \cdot \nabla \mathbf{u} = -\frac{1}{\rho} \nabla p + \nu \nabla^2 \mathbf{u},$$
$$\nabla \cdot \mathbf{u} = 0,$$
(3)

with $\nu = \mu / \rho$ being the kinematic viscosity. With the perturbation expansion (1), and using a time periodic expression for the first order field

$$\mathbf{u}^{(1)}(\mathbf{x},t) = \mathbf{u}_1(\mathbf{x})e^{i\omega t}, \qquad p^{(1)}(\mathbf{x},t) = p_1(\mathbf{x})e^{i\omega t}$$

one obtains for the momentum equation at $O(\varepsilon)$

$$i\omega \mathbf{u}_1 = -\frac{1}{\rho} \nabla p_1 + \nu \nabla^2 \mathbf{u}_1.$$
(4)

The equation (4) is solved using a Helmholtz decomposition

$$\mathbf{u}_1 = \mathbf{u}_\varphi + \mathbf{u}_A, \qquad \mathbf{u}_\varphi = \nabla \varphi, \qquad \nabla^2 \varphi = 0, \qquad \mathbf{u}_A = \nabla \times A, \qquad \nabla \cdot A = 0 \tag{5}$$

Note that the generality of the Nyborg formulation (Nyborg 1958) is premised on finding the velocity in terms of values of the potential component $\mathbf{u}_\varphi$ and its derivative at the boundary. The vortical part $\mathbf{u}_A$ satisfies

$$(\nabla^2 + h^2)\mathbf{u}_A, \qquad h^2 = -i\omega/\nu, \qquad h = (1-i)\sqrt{\frac{\omega}{2\nu}} = (1-i)\beta, \tag{6}$$

where the sign of $h$ was chosen for decaying solution of $\exp(-ihz)$. The solution has the typical structure of Stokes boundary layer for an oscillatory outer driving flow $\mathbf{u}_\varphi$ (of an order $\sim U$ and varying in a large scale $\sim L$) near a wall with boundary layer thickness $\delta = 2\pi/\beta \ll L$. We seek solution $(u, w)$ in an axisymmetric geometry. Accordingly, the solution of the radial component $u_A$ is straightforward and chosen to ensure a zero tangential velocity countering $u_\varphi$

$$u_A = -u_\varphi e^{-ihz}. \tag{7}$$

It satisfies (6). The axial component $w_A$ is chosen as

$$w_A = -\frac{\gamma}{ih} e^{-ihz}, \qquad \gamma = -\frac{\partial w_\varphi}{\partial z} = \frac{1}{r}\frac{\partial}{\partial r}(r u_\varphi) \tag{8}$$

Due to (5), $w_\varphi$ and $\gamma$ are harmonic, and therefore (8) satisfies (6). We note that

$$\nabla \cdot \mathbf{u}_\varphi = \frac{1}{r}\frac{\partial}{\partial r}(r u_\varphi) + \frac{\partial w_\varphi}{\partial z} = 0, \tag{9}$$

to obtain

$$\nabla \cdot \mathbf{u}_A = -\frac{1}{r}\frac{\partial}{\partial r}(r u_\varphi) e^{-ihz} - \frac{\partial w_\varphi}{\partial z} e^{-ihz} + \frac{1}{ih}\frac{\partial^2 w_\varphi}{\partial z^2} e^{-ihz} \approx 0,$$

the last term being higher order in the small quantity $\delta/L$ compared to the first two and therefore was neglected here as well as below (effectively $\gamma$ being treated as approximately a constant). The velocity $\mathbf{u}_\varphi + \mathbf{u}_A$ however does not satisfy the zero normal velocity condition at $z = 0$ due to $w_A$. Correcting for that a modified total first order velocity is found as

$$\mathbf{u} = \mathbf{u}_\varphi + \mathbf{u}_A + w_c \hat{\mathbf{e}}_z, \qquad w_c = \gamma/ih, \tag{10}$$

keeping in mind that the non-decaying $w_c$ is only meaningful while considering the velocity field in the small boundary layer region. Therefore, one obtains the time periodic first order velocity field (superscript [1] indicates the corresponding term with periodic time dependence included).

$$u^{(1)} = u_\varphi^{(1)} + u_A^{(1)} = u_\varphi [\cos \omega t - e^{-\beta z} \cos(\omega t - \beta z)],$$

$$w_A^{(1)} = w_\varphi^{(1)} + w_c^{(1)} + w_A^{(1)} = w_\varphi \cos \omega t + \frac{\gamma}{\sqrt{2}\beta}[\cos(\omega t - \pi/4) - e^{-\beta z} \cos(\omega t - \beta z - \pi/4)]. \quad (11)$$

This expression is consistent with Nyborg. At the second order, one observes the forcing term **F** in the right hand side of the average Stokes equation(2). As was noted by Nyborg, quadratic product of the irrotational part $\mathbf{u}_\varphi = \nabla \varphi$ does not contribute to streaming and can balance the pressure gradient term as in Bernoulli term

$$-\nabla p_2 = \mathbf{u}_\varphi \cdot \nabla \mathbf{u}_\varphi = \frac{1}{2}\nabla\left(|\nabla \varphi|^2\right), \quad (12)$$

reducing (2) into

$$\mu \nabla^2 <\mathbf{u}^{(2)}>_t = \mathbf{F} = \rho < (\mathbf{u}_A^{(1)} + w_c^{(1)}\hat{e}_z) \cdot \nabla \mathbf{u}_\varphi^{(1)} + \mathbf{u}^{(1)} \cdot \nabla (\mathbf{u}_A^{(1)} + w_c^{(1)}\hat{e}_z) >_t \quad (13)$$

Therefore, the forcing term in the $r$-direction becomes

$$F_r = \rho < \mathbf{u}_A^{(1)} \cdot \nabla u_\varphi^{(1)} + w_c^{(1)}\hat{e}_z \cdot \nabla u_\varphi^{(1)} + \mathbf{u}^{(1)} \cdot \nabla u_A^{(1)} >_t$$

$$= \rho < u_A^{(1)} \partial_r u_\varphi^{(1)} + w_A^{(1)} \partial_z u_\varphi^{(1)} + w_c^{(1)} \partial_z u_\varphi^{(1)} \quad (14)$$

$$+ u_\varphi^{(1)} \partial_r u_A^{(1)} + u_A^{(1)} \partial_r u_A^{(1)} + w_\varphi^{(1)} \partial_z u_A^{(1)} + w_c^{(1)} \partial_z u_A^{(1)} + w_A^{(1)} \partial_z u_A^{(1)} >_t$$

Noting the order of terms:

$$u_\varphi \sim u_A \sim w_\varphi \sim U,$$
$$w_A \sim w_c \sim \delta U/L,$$
$$\partial_r u_\varphi \sim \partial_z u_\varphi \sim \partial_r u_A \sim U/L,$$
$$\partial_z u_A \sim U/\delta,$$

and neglecting higher order in $\delta/L$, we obtain

$$F_r = \rho < u_A^{(1)} \partial_r u_\varphi^{(1)} + u_\varphi^{(1)} \partial_r u_A^{(1)} + u_A^{(1)} \partial_r u_A^{(1)} + w^{(1)} \partial_z u_A^{(1)} >_t. \quad (15)$$

Each term in the right-hand-side is $O(U^2/L)$. In an effort to express each term in terms of $u_\varphi$ and $\partial_r u_\varphi^{(1)}$, following Nyborg, we express the odd term $w_\varphi$ in (11) as

$$w_\varphi \approx z\left(\frac{\partial w_\varphi}{\partial z}\right)_{z=0} = -\frac{z}{r}\left[\frac{\partial}{\partial r}(r u_\varphi)\right]_{z=0}.$$

After substituting (11) in (15) averaging over time, one obtains

$$F_r = \frac{\rho}{2}\left(u_\varphi \partial_r u_\varphi (e^{-2\beta z} - 2e^{-\beta z}\cos\beta z) - u_\varphi (1/r)\partial_r (r u_\varphi) e^{-\beta z}\{\beta z(\cos\beta z + \sin\beta z) - \sin\beta z\}\right) \quad (16)$$

In the governing equation (13) in the second order, we note that the vertical ($z$) derivative is larger than the transverse ($r$) derivative by $O(L/\delta)$ to obtain

$$<\frac{\partial^2 u^{(2)}}{\partial^2 z}>_t = -\frac{F_r}{\mu},$$

Integrating and noting (8), we get

$$<u^{(2)}>_t = \frac{1}{\omega}\left(\frac{1}{2}\frac{\partial u_\varphi^2}{\partial r}(u_\alpha - u_\beta) - \frac{u_\varphi^2}{r}u_\beta\right), \quad \text{where}$$

$$u_\alpha = \frac{1}{4}e^{-2\beta z} + e^{-\beta z}\sin\beta z - \frac{1}{4}, \tag{17}$$

$$u_\beta = \frac{1}{2}\beta z e^{-\beta z}(\cos\beta z - \sin\beta z) - e^{-\beta z}\left(\sin\beta z + \frac{1}{2}\cos\beta z\right) + \frac{1}{2},$$

as was also found by Nyborg (1958).

The vertical component of streaming velocity $<w^{(2)}>_t$ is obtained by using equation of mass conservation and taking into account the no slip condition on the rigid wall:

$$<w^{(2)}>_t = \int_0^z -\frac{1}{r}\frac{\partial(r<u^{(2)}>)}{\partial r}dz$$

$$= -\frac{1}{2\omega}\left(\frac{1}{r}\frac{\partial u_\varphi^2}{\partial r}\right)\left(-\frac{1}{8\beta}e^{-2\beta z} - \frac{1}{4\beta}e^{-\beta z}(6\beta z\sin\beta z + 8\sin\beta z + 14\cos\beta z) - \frac{7}{4}z + \frac{29}{8\beta}\right)$$

$$-\frac{1}{2\omega}\left(\frac{\partial^2 u_\varphi^2}{\partial r^2}\right)\left(-\frac{1}{8\beta}e^{-2\beta z} - \frac{1}{4\beta}e^{-\beta z}(2\beta z\sin\beta z + 4\sin\beta z + 6\cos\beta z) - \frac{3}{4}z + \frac{13}{8\beta}\right).$$

(18)

Here $z$-dependence of $\partial u_\varphi^2/\partial r$ and $\partial^2 u_\varphi^2/\partial r^2$ are neglected in the boundary layer similar to what was assumed in (16).

The acoustic streaming velocity field is therefore known in terms of the outer irrotational velocity field. One can compute the shear stress on the wall:

$$\tau_{wall} = \mu\frac{\partial <u^{(2)}>_t}{\partial z}\bigg|_{z=0} = \frac{\rho_0}{4\beta}u_\varphi\frac{\partial u_\varphi}{\partial r}\bigg|_{z=0} \tag{19}$$

Doinikov and Boukaz described the microstreaming shear stress on the rigid wall assumed as a cell membrane due to spherical pulsating bubble detached from the wall (Doinikov and Bouakaz 2010).

## B. Potential velocity $u_\varphi$ due to oscillating bubble above a rigid surface:

In this research, we studied the microstreaming flow close to the plane rigid wall due to spherical pulsation of microbubbles in an ultrasound field. Figure 1 shows the schematic of the problem. The effect of the wall has been considered by assuming an image bubble at a distance 2h away from the real bubble which satisfies impermeability condition along radial axis.

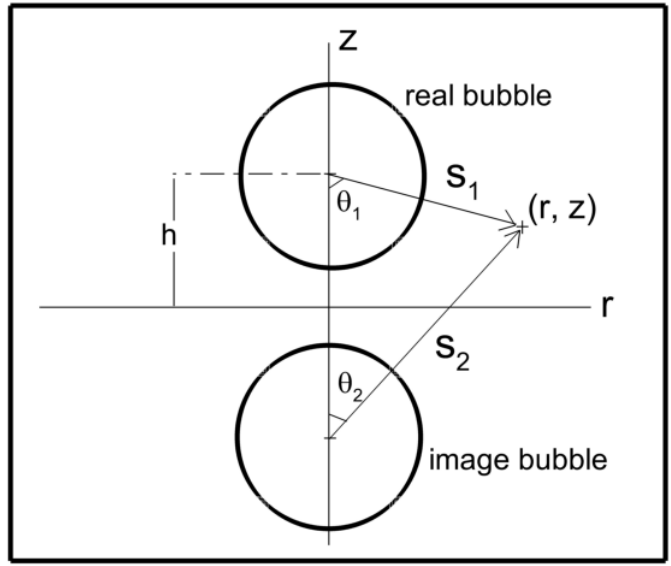

**Figure 1.** Schematic of the problem

As mentioned earlier, the streaming velocity can be found by having the local irrotational velocity distribution. The velocity potential $\phi$ of the fluid around the microbubble is:

$$\varphi = \left(\frac{1}{S_1} + \frac{1}{S_2}\right)\dot{R}R^2 . \tag{20}$$

$S_1$ and $S_2$ are the distances from the center of the real and image microbubbles to the desired location in the fluid, $\dot{R}$ and $R$ are the velocity and instant radius of the microbubble.

The irrotational local velocity $\nabla\phi$ components in the fluid particles in radial and vertical directions are as follows:

$$u^{(\varphi)} = u_\varphi e^{i\omega t} = \dot{R}R^2 \left( \frac{r}{\left(r^2+(z-h)^2\right)^{3/2}} + \frac{r}{\left(r^2+(z+h)^2\right)^{3/2}} \right),$$

(21)

$$w^{(\varphi)} = w_\varphi e^{i\omega t} = \dot{R}R^2 \left( \frac{z-h}{\left(r^2+(z-h)^2\right)^{3/2}} + \frac{z+h}{\left(r^2+(z+h)^2\right)^{3/2}} \right).$$

$u_\varphi$ and $w_\varphi$ are the amplitudes of potential velocity components in radial and vertical directions, $r$ and $z$ are the radial and vertical distance of the points in the fluid from origin and $h$ is the initial distance of the microbubble center from the wall.

The velocity $\dot{R}$ and the instant radius $R$ of the microbubble pulsating near the wall is described using the Rayleigh-Plesset type (R-P) equation shown in equation(22).

$$\rho\left(R\ddot{R} + \frac{3}{2}\dot{R}^2\right) = P_b - P_\infty,$$

(22)

Where,

$$P_b = P_{g_0}\left(\frac{R_0}{R}\right)^{3\kappa} - \frac{4\mu\dot{R}}{R} - \frac{2\gamma}{R} - P_{sc}(h,t),$$

(23)

$$P_{sc}(h,t) = \rho\frac{R}{2h}\left(R\ddot{R} + 2\dot{R}^2\right),$$

$$P_\infty = P_0 - P_{ex}\sin\omega t.$$

$P_b$ is the fluid pressure adjacent to the microbubble, $P_\infty$ is the pressure of the fluid at the far field, $\rho$ is the fluid density, $\mu$ is the fluid viscosity, $\gamma$ is the gas-fluid surface tension, $P_{g_0}$ is the initial gas pressure inside the microbubble, $P_0$ is the ambient pressure, $P_{ex}$ is the excitation pressure and $\kappa$ is the polytropic constant. Note that the effect of the wall has been considered as a pressure $P_{sc}(h,t)$ scattered from the image bubble located at a distance $2h$ from the real bubble. Since the microbubble is pulsating with small amplitude, equation (22) can be linearized to give analytical expression for the irrotational velocity of fluid particles.

## C. Linearization and normalizing:

For small pulsation $R = R_0(1+x)$ of the free bubble at distance $h$ from the wall, linearizing equation (22) in x and non-dimensionalizing $t$ and $P_{ex}$ with $t^* = t\omega$ and $P_{ex}^* = P_{ex}/P_0$ results in the non-dimensional equation of damped harmonic oscillator (equation(24)).

$$\ddot{x} + \left(\frac{4}{\mathcal{R}_e(1+R_0/2h)}\right)\dot{x} + \left(\frac{3\kappa \mathcal{E}_u}{1+R_0/2h} + \frac{6\kappa - 2}{\mathcal{W}_e(1+R_0/2h)}\right)x = \frac{P_{ex}^* \mathcal{E}_u \sin(\omega t)}{1+R_0/2h} , \quad (24)$$

where

$$\mathcal{R}_e = \frac{\rho R_0^2 \omega}{\mu}, \quad \mathcal{E}_u = \frac{P_0}{\rho R_0^2 \omega^2}, \quad \mathcal{W}_e = \frac{\rho R_0^3 \omega^2}{\gamma} , \quad (25)$$

are the characteristic Reynolds, Euler and Weber number of the microbubble.

$$\frac{\omega_0^2}{\omega^2} = \frac{3\kappa \mathcal{E}_u}{1+R_0/2h} + \frac{6\kappa - 2}{\mathcal{W}_e(1+R_0/2h)},$$

$$\delta_t = \frac{4\omega}{\mathcal{R}_e(1+R_0/2h)\omega_0} \quad (26)$$

$\omega_0$ and $\delta_t$ are the linear angular natural frequency and damping term of the microbubble. The analytic solution of equation (24) in steady region when the transient region has been subsided is as follows:

$$x = \frac{\mathcal{E}_u P_{ex}^* \sin(\omega t + \phi)}{Z_m(1+R_0/2h)} = \xi_m \sin(\omega t + \phi) , \quad (27)$$

where $Z_m = \sqrt{(\delta_t \omega_0/\omega)^2 + (1/\omega^4)(\omega_0^2 - \omega^2)^2}$ is the non-dimensional absolute value of impedance and $\phi$ is the oscillation phase.

For small pulsation of microbubble, the radial component of the amplitude of linearized non-dimensional potential velocity $u_\varphi^*$ and non-dimensional radial streaming velocity $<u^{(2)}>_t^*$ (with respect to $R_0 \omega$) are in the following form:

$$u_\varphi^* = \xi_m \left( \frac{r/R_0}{\left((r/R_0)^2 + (z/R_0 - h/R_0)^2\right)^{3/2}} + \frac{r/R_0}{\left((r/R_0)^2 + (z/R_0 + h/R_0)^2\right)^{3/2}} \right), \quad (28)$$

$$<u^{(2)}>_t^* = \frac{1}{2}\frac{\partial u_\varphi^{*2}}{\partial r^*}(u_\alpha - u_\beta) - \frac{u_\varphi^{*2}}{r^*}u_\beta ,  \qquad (29)$$

where $r^* = r/R_0$, $u_\alpha$ and $u_\beta$ are calculated from (17).

The linearized form of shear stress (19) in non-dimensional form (with respect to ambient pressure) is as follows:

$$\tau^* = \frac{1}{\mathcal{E}_u(2\mathcal{R}e)^{1/2}}\xi_m^2\left(\frac{R_0}{h}\right)^5\left[\frac{2(r/h)(1-2(r/h)^2)}{(1+(r/h)^2)^4}\right] \qquad (30)$$

## 3. Results

### A. *Microstreaming due to free bubble oscillation:*

The microbubble has been excited with ultrasound wave at excitation frequency $f_{ex}$ and excitation pressure $P_{ex}$. The initial radius of the microbubble is $R_0$.

Figure 2 shows the radial pulsation of the microbubble with non-dimensional time located at $h = 2R_0$ when $\mathcal{R}e = 85, \mathcal{E}_u = 0.125, \mathcal{W}e = 33$ (corresponding to $P_{ex} = 100Kpa, f_{ex} = 1.5MHz, R_0 = 3\mu m$). As it is shown in the steady region, the maximum pulsation amplitude of the microbubble is $0.17R_0$. The same pulsation amplitude is obtained using linearized R-P equation (equation(24)).

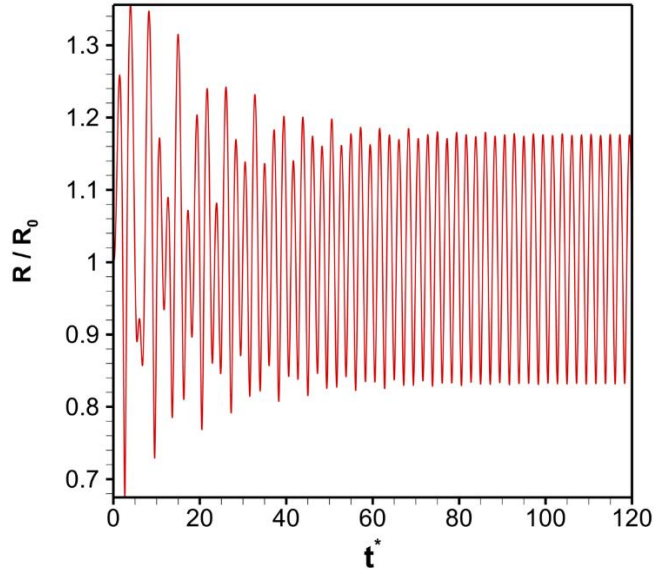

**Figure 2.** The radial oscillation of the free microbubble when

$$\mathcal{R}e = 85, \mathcal{E}_u = 0.125, \mathcal{W}e = 33, h/R_0 = 2.0$$

As it is shown in equation(2), the temporal average of $\mathbf{u}^{(1)} \cdot \nabla \mathbf{u}^{(1)}$ acts as an external force which drives the streaming motion, where $\mathbf{u}^{(1)}$ is the first order sinusoidal velocity which can be written as the function of potential velocity (equation(11)). Figure 3 shows the potential velocity surrounding the bubble at two different non-dimensional times during microbubble expansion and compression.

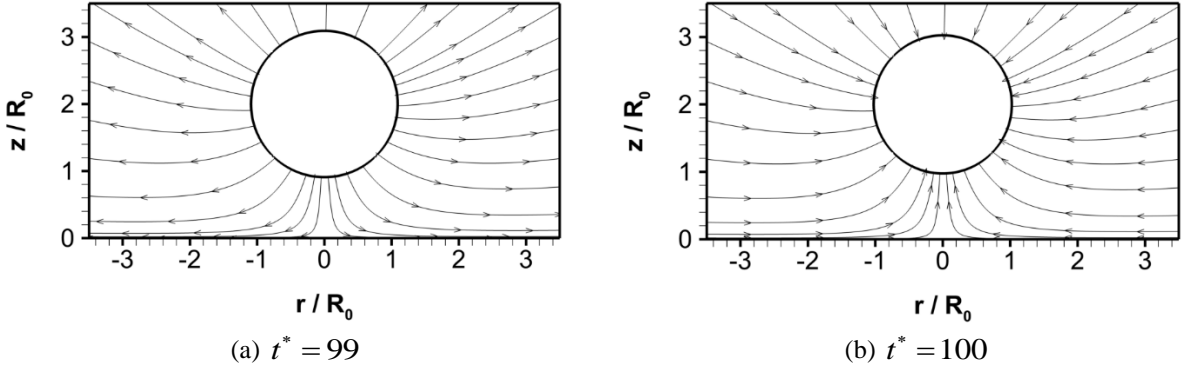

(a) $t^* = 99$  (b) $t^* = 100$

**Figure 3.** The irrotational velocity around the free microbubble when $\mathcal{R}e = 85, \mathcal{E}u = 0.125, \mathcal{W}e = 33, h/R_0 = 2.0$

(a) during expansion, (b) during compression

Figure 4 shows the amplitude of the radial component of potential velocity along the rigid wall obtained from equation(28). It is shown that for a microbubble with $\mathcal{R}e = 85, \mathcal{E}u = 0.125, \mathcal{W}e = 33$ located at $h = 2R_0$, the amplitude of radial potential velocity on the wall has a maximum peak at $r = 1.41R_0$. Figure 4 states that $\partial u_\varphi^* / \partial r^*$ and hence $< u^{(1)} \partial u^{(1)} / \partial r >_t$ is positive near the wall for radial distances $r < 1.41R_0$, and it is negative for radial distances $r > 1.41R_0$. Therefore, the direction of the radial external force driving the microstreaming flow $\rho_0 < u^{(1)} \cdot \nabla u^{(1)} >_t$ is changing at $r = 1.41R_0$. The change in external force direction pushes the fluid upward (as the flow cannot go downward because of the wall) and creates vortical motion.

To observe the microstreaming flow, figure 5(a) shows the streamlines due to microstreaming near the wall for the condition described in figure 3.

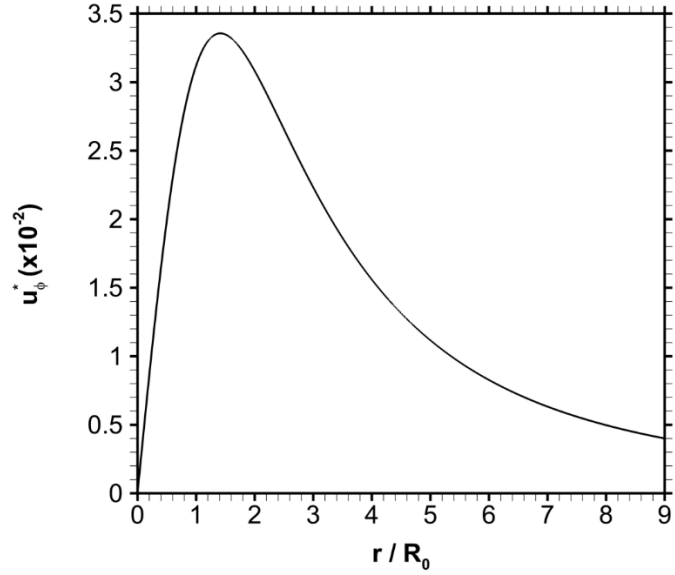

**Figure 4.** The amplitude of radial potential velocity on the rigid wall when
$\mathcal{R}e = 85, \mathcal{E}u = 0.125, \mathcal{W}e = 33, h/R_0 = 2.0$

As it is observed in figure 5(a), an axisymmetric vortex is generated near the wall with the length of $L = 1.41 R_0$ corresponding to the location where $u_\varphi^*$ is maximum or the location where the direction of external force driving streaming flow changes. Inside the vortex, the flow near the wall is directed inward, while it is directed radially outward beyond $r = 1.41 R_0$. Figure 5(b) shows the shear stress on the wall due to microstreaming when transient pulsation of microbubble has been subsided. As it is expected the sign of the shear stress changes at $r = 1.41 R_0$ corresponding to the location where the direction of the streaming flow near the wall changes.

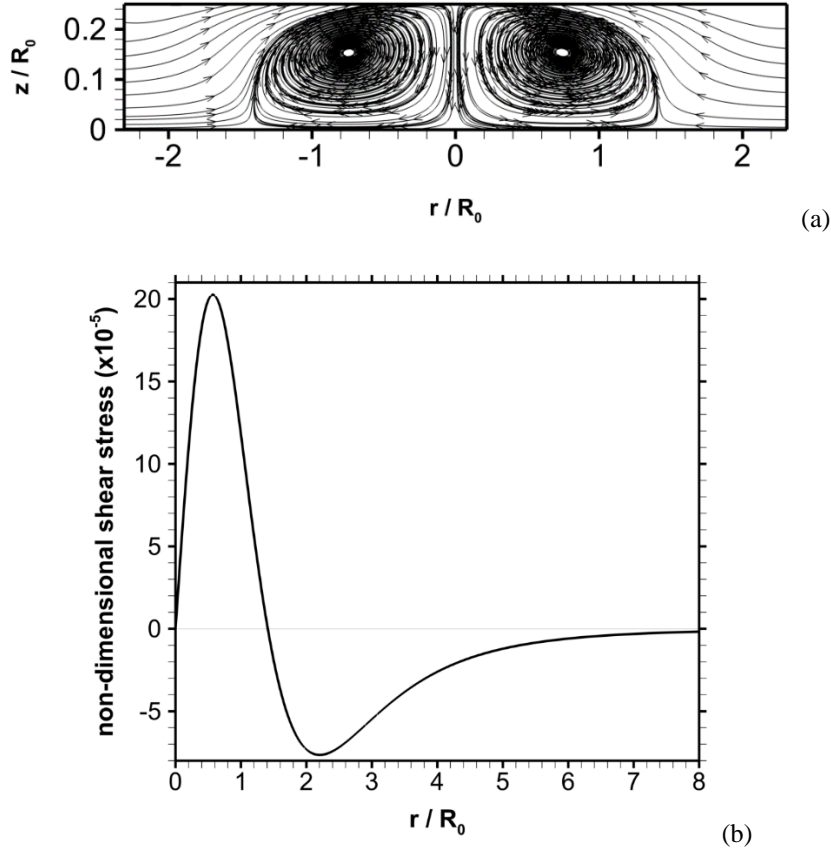

**Figure 5.** (a) Microstreaming streamlines near the plane rigid wall due to free microbubble, (b) the induced shear stress on the wall when $\mathcal{R}e = 85, \mathcal{E}u = 0.125, \mathcal{W}e = 33, h/R_0 = 2.0$

As mentioned above, the length of the vortex can be determined when the direction of radial external force $<u^{(1)} \partial u^{(1)}/\partial r>_t$ driving the microstreaming is changing corresponding to the location where radial potential velocity on the wall is maximum or the location on the wall where the shear stress becomes zero. Therefore:

$$\left.\frac{\partial u_\varphi^*}{\partial r^*}\right|_{z^*=0} \rightarrow \quad \frac{L}{h} = \frac{\sqrt{2}}{2}, \tag{31}$$

where $L$ is the vortex length. As it is observed in equation(31), for a spherically pulsating bubble, the length of the vortex is only dependent on the initial distance of the microbubble from the rigid wall $h$.

Figure 6(a-b) shows the streamlines near the plane wall due to microstreaming and the induced shear stress on the wall when $\mathcal{R}e = 85, \mathcal{E}u = 0.125, \mathcal{W}e = 33, h/R_0 = 3.0$. It is seen that the maximum shear stress on the wall decreased as the bubble excited further away from the wall.

Also the length of the vortex increased to $L = 2.12 R_0$ as the microbubble moves away from the wall (figure 6(a)). The increase of the length of the vortex can be explained in figure 7 which shows the non-dimensional potential velocity amplitude along the radial distance on the wall. Comparing figure 7 with figure 4 shows that the maximum amplitude of the potential velocity shifts right when the bubble is excited further away from the wall. This shift states that when the microbubble is excited further away from the wall, the direction of the radial driving force changes at radial distances further far from the origin.

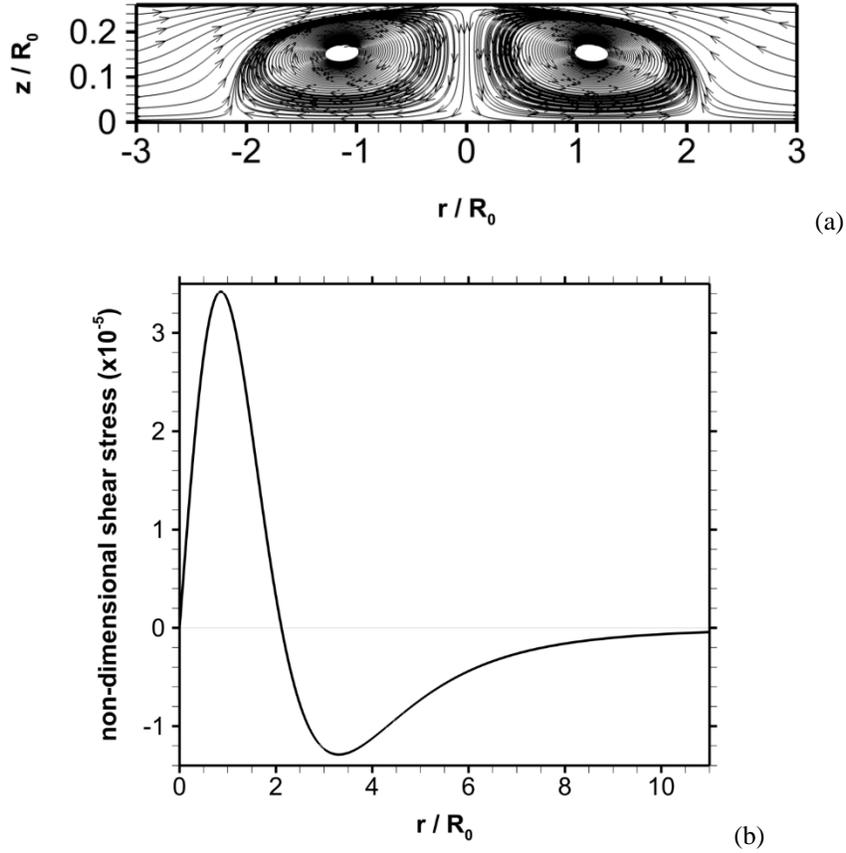

**Figure 6.** (a) Microstreaming streamlines near the plane rigid wall due to free microbubble, (b) the induced shear stress on the wall when $\mathcal{R}e = 85, \mathcal{E}u = 0.125, \mathcal{W}e = 33, h/R_0 = 3.0$

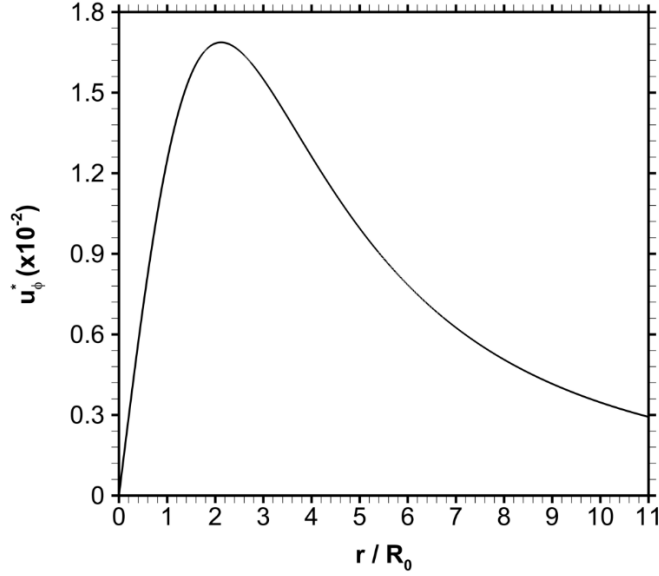

**Figure 7.** The amplitude of radial potential velocity on the rigid wall when
$\mathcal{R}e = 85, \mathcal{E}u = 0.125, \mathcal{W}e = 33, h/R_0 = 3.0$

## B. *Effect of translational motion of microbubble on microstreaming:*

The microbubble tends to translate toward the wall due to Bjerkness force when it is pulsating. For a microbubble with both pulsation and translation, the velocity potential in the surrounding fluid is as follows:

$$\varphi = \left(\frac{1}{S_1} + \frac{1}{S_2}\right)\dot{R}R^2 + b\left(\frac{1}{S_1}\right)^2(\cos\theta_1) + b\left(\frac{1}{S_2}\right)^2(\cos\theta_2), \tag{32}$$

where $\theta_1$ and $\theta_2$ are shown in figure 1. The amplitude of linearized non-dimensional potential velocity in radial direction will be in the following form:

$$u_\varphi^* = \xi_m \left( \frac{r/R_0}{\left((r/R_0)^2 + (h/R_0 - z/R_0)^2\right)^{3/2}} + \frac{r/R_0}{\left((r/R_0)^2 + (h/R_0 + z/R_0)^2\right)^{3/2}} \right)$$
$$+ b^t \left( \frac{3(h/R_0 - z/R_0)r/R_0}{\left((r/R_0)^2 + (h/R_0 - z/R_0)^2\right)^{5/2}} + \frac{3(h/R_0 + z/R_0)r/R_0}{\left((r/R_0)^2 + (h/R_0 + z/R_0)^2\right)^{5/2}} \right), \tag{33}$$

where $b^t$ is as follows (Doinikov and Bouakaz 2014):

$$b^I = \left| \xi_m \left( \frac{R_0}{2h} \right)^2 \left( \frac{\alpha^3 + 3i\alpha^2 - 6\alpha - 6i}{-\alpha^3 - 3i\alpha^2 + 18\alpha + 18i} \right) \right|, \tag{34}$$

$$\alpha = (1+i)R_0\beta .$$

The non-dimensional potential velocity in (33) is used to calculate the non-dimensional microstreaming velocity (equation(29)). Figure 8 shows the comparison of radial streaming velocity at $r = 0.5R_0$ for the two cases when the translational motion of the microbubble has been considered and ignored. It is observed that for the conditions mentioned in this study, the translational motion has no considerable effect in the microstreaming flow, and hence it is neglected in our study.

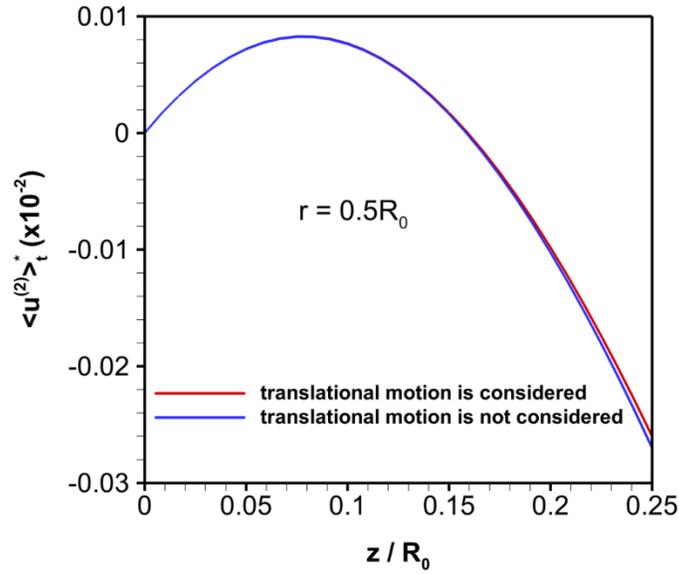

**Figure 8.** The radial streaming velocity when $\mathcal{R}e = 85, \mathcal{E}u = 0.125, \mathcal{W}e = 33, h/R_0 = 2.0$

### C. Microstreaming due to oscillation of coated microbubble:

#### C.1. Viscoelastic Strain-softening Exponential Elasticity Model (EEM) for encapsulation:

To study the microstreaming due to coated microbubbles, we have studied Sonazoid contrast agent. Contrast agents have been initially developed for enhancing the contrast of the image in ultrasound imaging. Sonazoid contrast agents are gas core microbubbles coated with a layer of lipid to stabilize them against early dissolution in the blood stream. To model the coating of these contrast agents, we have used an interfacial rheology model called exponential elasticity model (EEM)(Paul, et al. 2010), where the coating is assumed to be a viscoelastic interface having dilatational viscosity and shell elasticity incorporating the effects of strain softening by letting the interfacial elasticity decrease exponentially with increasing surface area.

Using EEM, the effective interfacial tension would be in the following form:

$$\gamma_{eff}(R) = \gamma_0 + \beta^s E^s, \qquad (35)$$

where $E^s = E_0^s \exp(-\alpha^s \beta^s)$ and $\beta^s = \left(\dfrac{R}{R_E}\right)^2 - 1$.

$\gamma_{eff}(R)$ and $E^s$ are the effective surface tension and the shell elasticity. $R_E = R_0 \left[1 + \left(1 - \sqrt{1 + 4\gamma_0 \alpha^s / E_0^s}\right)/2\alpha^s\right]^{-1/2}$ is the equilibrium radius of the contrast agent. $\gamma_0$, $\alpha^s$, $E_0^s$ are constants dependent to physical properties of the coating. Due to the coating, the fluid pressure on the microbubble would be as follows:

$$P_b = P_{g_0}\left(\dfrac{R_0}{R}\right)^{3\kappa} - \dfrac{4\kappa^s \dot{R}}{R^2} - \dfrac{4\mu \dot{R}}{R} - \dfrac{2\gamma_{eff}}{R} - P_{sc}(h,t). \qquad (36)$$

$\kappa^s$ is the dilatational viscosity due to the bubble coating. The characteristic properties of Sonazoid according to exponential elasticity model (EEM) are as follows (Katiyar and Sarkar 2011):

$$\gamma_0 = 0.019\, N/m,\; E_0^s = 0.55\, N/m,\; \alpha^s = 1.5,\; \kappa^s = 1.2 \times 10^{-8}\, Ns/m$$

Substituting $P_b$ from equation (36) in equation (22) gives the modified Rayleigh-Plesset type equation for pulsation of coated microbubbles.

## C.2 Linearization and normalizing of pulsation of coated microbubble:

Again, for small pulsations, the Rayleigh-Plesset equation (R-P) can be linearized in x where $R = R_0(1+x)$ to obtain the damped harmonic oscillator.

For a coated microbubble near a rigid wall, linearizing R-P equation using EEM for coating gives the following linear angular natural frequency $\omega_0$ and damping term $\delta_t$. Substituting (37) in (27) gives the analytic solution of the non-dimensional displacement x of coated microbubble. By having x, the non-dimensional streaming velocity and non-dimensional shear stress due to contrast microbubble can be calculated analytically through (28)-(30).

$$\dfrac{\omega_0^2}{\omega^2} = \dfrac{3\kappa \mathcal{E}_u}{1 + R_0/2h} + \dfrac{2\left(\sqrt{1 + 4\gamma_0 \alpha^s / E_0^s}/\alpha^s\right)\left(1 + 2\alpha^s - \sqrt{1 + 4\gamma_0 \alpha^s / E_0^s}\right)}{\mathcal{W}_e^s (1 + R_0/2h)},$$

$$\delta_t = \dfrac{4\omega}{(1 + R_0/2h)\omega_0}\left(\dfrac{1}{\mathcal{R}_e} + \dfrac{1}{\mathcal{R}_e^s}\right), \qquad (37)$$

where $\mathcal{W}_e^s = \dfrac{\rho R_0^3 \omega^2}{E_0^s}$, $\mathcal{R}_e^s = \dfrac{\rho R_0^3 \omega}{\kappa^s}$ and $\mathcal{E}_u$, $\mathcal{W}_e$, $\mathcal{R}_e$ are calculated from (25).

Figure 9 shows the streamlines of microstreaming flow near the wall due to pulsation of Sonazoid located initially at $h = 2.5R_0$ when $\mathcal{R}e = 85, \mathcal{E}u = 0.031, \mathcal{W}e = 33, \mathcal{R}e^s = 21, \mathcal{W}e^s = 4$ (corresponding to $P_{ex} = 25Kpa, f_{ex} = 1.5MHz, R_0 = 3\mu m$). The streamlines look very similar to the case when the bubble has no coating. As it is seen in figure 9, the length of the vortex is $1.77R_0$ corresponding to the vortex length due to free microbubble located at the same distance from the wall (equation (31) ).

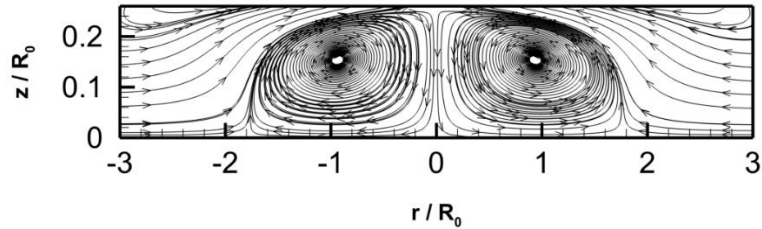

**Figure 9.** Streaming velocity near the plane rigid wall due to coated microbubble when
$\mathcal{R}e = 85, \mathcal{E}u = 0.031, \mathcal{W}e = 33, \mathcal{R}e^s = 21, \mathcal{W}e^s = 4, h/R_0 = 2.5$

**Conclusion**

In this study, the acoustic microstreaming flow near a plane rigid wall has been studied. Microbubbles are assumed to pulsate spherically near the wall under the excitation of low amplitude ultrasound. It has been shown that an axisymmetric vortex generated near the wall. The vortex is being generated due to the change in the direction of microstreaming driving force near the wall. Near the wall, the driving force is related to the gradient of the potential velocity of the fluid. Potential velocity on the wall has a maximum peak causing the change in the direction of microstreaming driving force. The length of the vortex has been shown to depend only on the distance of the microbubble from the rigid wall whether the microbubble is free (uncoated) or coated.